\documentclass[a4paper,fleqn,usenatbib]{mnras}

\usepackage{newtxtext,newtxmath}

\usepackage[T1]{fontenc}
\usepackage{ae,aecompl}
\usepackage{graphicx}   
\usepackage{amsmath}    
\usepackage{amsfonts}
\usepackage{bm,color}

\title{$R^p$ Attractors Static Neutron Star Phenomenology}
\author[Oikonomou]{Vasilis K. Oikonomou$^{1,2}$\\
\small $^{1}$ Department of Physics, Aristotle University of
Thessaloniki, Thessaloniki 54124,
Greece \\
\small $^{2}$ Institut f\"{u}r Theoretische Physik, Goethe
Universit\"{a}t Frankfurt, Max-von-Laue-Str.1, 60438 Frankfurt am
Main, Germany }

\begin{document}
\label{firstpage}
\pagerange{\pageref{firstpage}--\pageref{lastpage}} \maketitle

\maketitle

\begin{abstract}
In this work we study the neutron star phenomenology of $R^p$
attractor theories in the Einstein frame. The Einstein frame $R^p$
attractor theories have the attractor property that they originate
from a large class of Jordan frame scalar theories with arbitrary
non-minimal coupling. These theories in the Einstein frame provide
a viable class of inflationary models, and in this work we
investigate their implications on static neutron stars. We
numerically solve the Tolman-Oppenheimer-Volkoff equations in the
Einstein frame, for three distinct equations of state, and we
provide the mass-radius diagrams for several cases of interest of
the $R^p$ attractor theories. We confront the results with several
timely constraints on the radii of specific mass neutron stars,
and as we show, only a few cases corresponding to specific
equations of state pass the stringent tests on neutron stars
phenomenology.
\end{abstract}

\begin{keywords}
stars: neutron; Physical Data and Processes, cosmology: theory
\end{keywords}

\section*{Introduction}

The direct gravitational wave observation GW170817
\cite{TheLIGOScientific:2017qsa,Abbott:2020khf} initiated what is
nowadays known as gravitational wave astronomy. Neutron stars (NS)
\cite{Haensel:2007yy,Friedman:2013xza,Baym:2017whm,Lattimer:2004pg,Olmo:2019flu}
are at the core of astrophysical gravitational wave observations,
and numerous scientific areas are jointly studying NS from their
perspective, for example nuclear theory
\cite{Lattimer:2012nd,Steiner:2011ft,Horowitz:2005zb,Watanabe:2000rj,Shen:1998gq,Xu:2009vi,Hebeler:2013nza,Mendoza-Temis:2014mja,Ho:2014pta,Kanakis-Pegios:2020kzp,Tsaloukidis:2022rus},
high energy physics
\cite{Buschmann:2019pfp,Safdi:2018oeu,Hook:2018iia,Edwards:2020afl,Nurmi:2021xds},
modified gravity
\cite{Astashenok:2020qds,Astashenok:2021peo,Capozziello:2015yza,Astashenok:2014nua,Astashenok:2014pua,Astashenok:2013vza,Arapoglu:2010rz,Panotopoulos:2021sbf,Lobato:2020fxt,Numajiri:2021nsc}
and astrophysics
\cite{Altiparmak:2022bke,Bauswein:2020kor,Vretinaris:2019spn,Bauswein:2020aag,Bauswein:2017vtn,Most:2018hfd,Rezzolla:2017aly,Nathanail:2021tay,Koppel:2019pys,Raaijmakers:2021uju,Most:2020exl,Ecker:2022dlg,Jiang:2022tps}.
The perspective of modified gravity implications on NS has been
for a long time in the mainstream of NS works, see for example
\cite{Astashenok:2014nua,Astashenok:2014pua} and also Refs.
\cite{Pani:2014jra,Staykov:2014mwa,Horbatsch:2015bua,Silva:2014fca,Doneva:2013qva,Xu:2020vbs,Salgado:1998sg,Shibata:2013pra,Arapoglu:2019mun,Ramazanoglu:2016kul,AltahaMotahar:2019ekm,Chew:2019lsa,Blazquez-Salcedo:2020ibb,Motahar:2017blm,Odintsov:2021qbq,Odintsov:2021nqa,Oikonomou:2021iid,Pretel:2022rwx,Pretel:2022plg,Cuzinatto:2016ehv}
for scalar-tensor descriptions of NS phenomenology. The main
effect of modified gravity descriptions of NS is the significant
elevation of the maximum NS masses, with modified gravity bringing
this maximum mass near or inside the mass-gap region with $M \geq
2.5 \,M_{\odot}$. Regarding non-minimally coupled scalar field
theories, there exists a vast class of viable inflationary
potentials which have the remarkable property of being attractors
\cite{alpha0,alpha1,alpha2,alpha3,alpha4,alpha5,alpha6,alpha7,alpha7a,alpha8,alpha9,alpha10,alpha11,alpha12,alpha13,alpha14,alpha15,alpha16,alpha17,alpha18,alpha19,alpha20,alpha21,alpha22,alpha23,alpha24,alpha25,alpha26,alpha27,alpha28,alpha29,alpha30,alpha31,alpha32,alpha33,alpha34,alpha35,alpha36,alpha37}.
The attractor terminology is justified due to the fact that
distinct non-minimally coupled scalar-tensor inflationary
theories, lead to the same Einstein frame inflationary
phenomenology, which is compatible with the latest Planck data
\cite{Akrami:2018odb}. The question always when studying these
attractor models is whether these models can be distinguished in
some way, phenomenologically. From an inflationary point of view,
and regarding the large wavelength Cosmic Microwave Background
modes, a discrimination between these models is impossible.
However, this discrimination is possible if NS are studied.
Indeed, the phenomenologically indistinguishable attractor models
can be discriminated in NS and vice versa, with the latter feature
being phenomenal. That is, if some models are indistinguishable
with respect to their NS phenomenology, they can be distinguished
if their inflationary properties are studied. To address these
issues in a concrete way, in this work we shall study $R^p$
attractor theories. The inflationary phenomenology of these
theories is studied in the recent literature \cite{rpattractors}
see also \cite{Motohashi:2014tra,Renzi:2019ewp} for subcases of
the original $R^p$ attractors theories. For a spherically
symmetric metric we derive and solve numerically the Einstein
frame Tolman-Oppenheimer-Volkoff (TOV) equations, using an LSODA
based double shooting python 3 numerical integration
\cite{niksterg}. We derive the Jordan frame $M-R$ graphs for the
$R^p$ attractors, for three different piecewise polytropic
\cite{Read:2008iy,Read:2009yp} equations of state (EoS), WFF1
\cite{Wiringa:1988tp}, the SLy \cite{Douchin:2001sv}, and the APR
EoS \cite{Akmal:1998cf}, using the Arnowitt-Deser-Misner (ADM)
definition of Jordan frame masses of NS \cite{Arnowitt:1960zzc}.
The NSs temperature is significantly lower than the Fermi energy
of the constituent particles of NSs, thus NS matter can be in
principle described by a single-parameter EoS that may describe
perfectly cold matter at densities higher than the nuclear
density. However, a serious problem emerges, having to do with the
uncertainty in the EoS, which is larger, and the pressure as a
function of the baryonic mass density cannot be accurately defined
and is uncertain to one order of magnitude at least above the
nuclear density. Moreover, the exact nature of the phase of matter
at the NSs core is highly uncertain. Hence, a parameterized-type
EoS at high densities is an optimal choice for an EoS, thus
rendering the piecewise polytropic EoS a suitable choice. In order
to construct the piecewise polytropic EoS, astrophysical
constraints are taken into account, both observational and
theoretical, like the causality constraints, see
\cite{Read:2008iy,Read:2009yp}, to also confirm the causality
fulfilment for all the piecewise polytropic EoS we shall use in
this paper. For the construction of the piecewise polytropic EoS
one uses a low-density part with $\rho<\rho_0$, which is basically
chosen to be a tabulated and well-known EoS for the crust, and
furthermore, the piecewise polytropic EoS also has a large density
part with $\rho\gg \rho_0$. We finally confront the resulting NS
phenomenologies with several recent constraints on the radii of
specific mass NS
\cite{Altiparmak:2022bke,Raaijmakers:2021uju,Bauswein:2017vtn} and
as we show, only a few scenarios and EoS are compatible with the
constraints on NS radii. Obviously, the gravitational wave
astronomy era has changed the way of thinking on theoretical
astrophysics, since several models of scalar-tensor gravity which
in the recent past could be considered as viable, nowadays may no
longer be valid.

\section{Inflationary Phenomenology of $R^p$ Attractors}

The full analysis of the generalized $R^p$ attractors is given in
Ref. \cite{rpattractors}, so we refer the reader for details. Here
we shall briefly discuss the inflationary phenomenological
properties of $R^p$ attractors in order to stress their importance
among other cosmological attractors
\cite{alpha0,alpha1,alpha2,alpha3,alpha4,alpha5,alpha6,alpha7,alpha7a,alpha8,alpha9,alpha10,alpha11,alpha12,alpha13,alpha14,alpha15,alpha16,alpha17,alpha18,alpha19,alpha20,alpha21,alpha22,alpha23,alpha24,alpha25,alpha26,alpha27,alpha28,alpha29,alpha30,alpha31,alpha32,alpha33,alpha34,alpha35,alpha36,alpha37}.
The $R^p$ attractors constitute a class of their own among other
attractors, and all the $R^p$ attractors in the Einstein frame
correspond to generalizations of the following Einstein frame
potential,
\begin{equation}\label{rpinfini}
V(\varphi)= V_0\,M_p^4e^{-2\sqrt{\frac{2}{3}}\kappa
\varphi}\left(e^{\sqrt{\frac{2}{3}}\kappa \varphi}
-1\right)^{\frac{p}{p-1}}\, ,
\end{equation}
where $M_p=\frac{1}{\sqrt{8\pi G}}$ is the reduced Planck mass and
$G$ is Newton's gravitational constant. The inflationary
properties of the above theory have been addressed in the recent
literature, see for example
\cite{Motohashi:2014tra,Renzi:2019ewp}. The scalar-tensor theory
with the potential (\ref{rpinfini}) corresponds to the Jordan
frame $F(R)$ gravity,
\begin{equation}\label{JordanframeFR}
F(R)=R+\beta R^p\, ,
\end{equation}
with $\beta$ is a free parameter with its physical dimensions in
natural units being $[\beta ]=[m]^{2-2p}$. The $R^p$ attractors
have the following scalar potential in the Einstein frame,
\begin{equation}\label{rpinfinialpha}
V(\varphi)= V_0\,M_p^4e^{-2\sqrt{\frac{2}{3 \alpha}}\kappa
\varphi}\left(e^{\sqrt{\frac{2}{3\alpha }}\kappa \varphi}
-1\right)^{\frac{p}{p-1}}\, ,
\end{equation}
where $M_p$ is the reduced Planck mass, and for $\alpha=1$ we
obtain the scalar theory with scalar potential
(\ref{rpinfinialpha}). Now the question is why these models are
classified as attractor models, what justifies the terminology
attractors? It is the class of scalar-tensor Jordan frame theories
which correspond to the Einstein frame potential
(\ref{rpinfinialpha}) that justify the use of the terminology
attractors. Basically, the potential (\ref{rpinfinialpha}) can be
the Einstein frame potential for a large class of Jordan frame
scalar-tensor theories, as we now evince. The $\phi$-Jordan frame
action is,
\begin{equation}\label{phijordanframeaction}
\mathcal{S}_J=\int
d^4x\left(\frac{\Omega(\phi)}{2\kappa^2}R-\frac{\omega
(\phi)}{2}g^{\mu \nu} \partial_{\mu}\phi \partial_{\nu}\phi
-V_J(\phi)\right)\, ,
\end{equation}
with the scalar field describing a non-canonical scalar field in
the Jordan frame, and the coupling function has the general form
$\Omega(\phi)=1+\xi f(\phi)$ with $\xi$ and $f(\phi)$ being the
arbitrary dimensionless coupling and an arbitrary dimensionless
function respectively. The $R^p$ attractors have the following
$\phi$-Jordan frame scalar potential,
\begin{equation}\label{potentialJordanframe}
V_J(\phi)=V_0\left(\Omega(\phi)-1 \right)^{\frac{p}{p-1}}\, ,
\end{equation}
and more importantly, the kinetic term function $\omega (\phi)$
has the following form,
\begin{equation}\label{kinetictermfunction}
\omega (\phi)=\frac{1}{4\xi}\frac{\left(\frac{d \Omega (\phi)}{d
\phi}\right)^2}{\Omega(\phi)}\, .
\end{equation}
Hence the large class of the $R^p$-attractors correspond to the
Jordan frame theories which are described by Eqs.
(\ref{potentialJordanframe}) and (\ref{kinetictermfunction}).
Notice that the Jordan frame functions $f(\phi)$ are arbitrary and
we shall not need to specify these. By performing the conformal
transformation of the Jordan frame metric $g_{\mu \nu}$,
\begin{equation}\label{conformaltransmetr}
\tilde{g}_{\mu \nu}=\Omega(\phi)g_{\mu \nu}\, ,
\end{equation}
we get the Einstein frame action,
\begin{equation}\label{alphaact}
\mathcal{S}_E=\sqrt{-\tilde{g}}\left(\frac{\tilde{R}}{2\kappa^2}-\tilde{g}^{\mu
\nu}\partial_{\mu}\varphi
\partial_{\nu}\varphi-V(\varphi)\right)\,
,
\end{equation}
with $\tilde{g}_{\mu \nu}$ denoting the Einstein frame metric
tensor, and the ``tilde'' indicates Einstein frame quantities.
Also the Einstein frame potential $V(\phi)$ and the Jordan frame
potential $V_J(\phi)$ are related as follows,
\begin{equation}\label{jordaneinsteinframepotential}
V(\varphi)=\Omega^{-2}(\phi)V_J(\phi)\, .
\end{equation}
Notice that the general relation which connects the Jordan frame
scalar field $\phi$ with the canonical Einstein frame scalar field
$\varphi$ is,
\begin{equation}\label{generalrelationbetweenscalar}
\left( \frac{d \varphi}{d
\phi}\right)^{2}=\frac{3}{2}\frac{\left(\frac{d \Omega (\phi)}{d
\phi}\right)^2}{\Omega(\phi)}+\frac{\omega (\phi)}{\Omega(\phi)}\,
,
\end{equation}
hence for the $R^p$ attractors, in which case the kinetic term
function $\omega (\phi)$ is chosen to be that of Eq.
(\ref{kinetictermfunction}), we finally have the important
relation of the non-minimal scalar coupling function to gravity,
\begin{equation}\label{omegaphigeneral}
\Omega (\phi)=e^{\sqrt{\frac{2}{3\alpha}}\varphi}\, ,
\end{equation}
with the parameter $\alpha$ being defined to be,
\begin{equation}\label{alphageneraldefinition}
\alpha=1+\frac{1}{6\xi}\, .
\end{equation}
Notice that by substituting Eq. (\ref{omegaphigeneral}) in Eq.
(\ref{jordaneinsteinframepotential}) we obtain the generalized
$R^p$-attractor potential of Eq. (\ref{rpinfinialpha}).
\begin{figure}
\centering
\includegraphics[width=20pc]{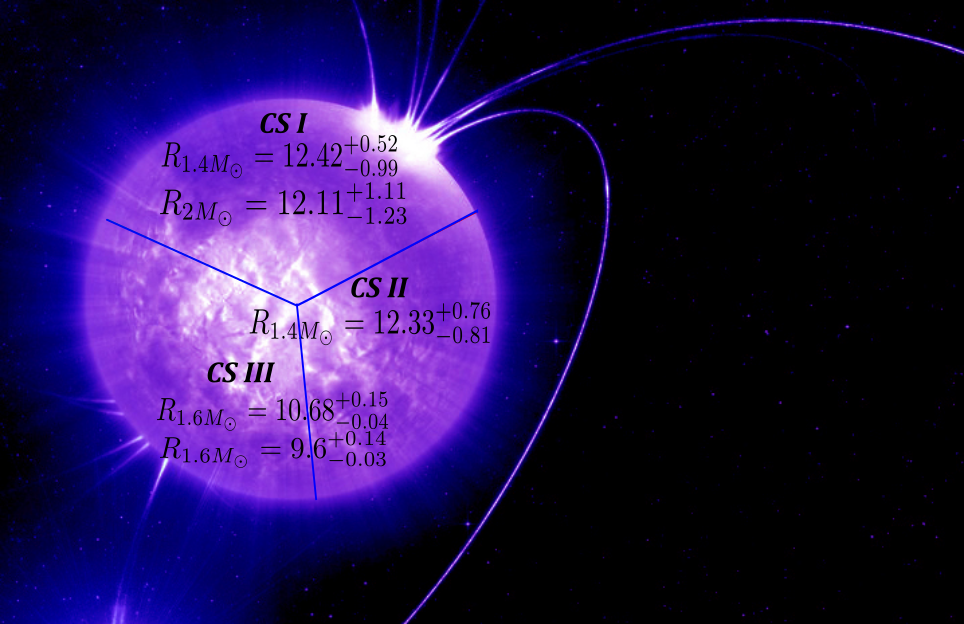}
\caption{The constraints CSI, CSII
 and CSIII. This figure is inspired and based after editing on Credit: ESO/L.Cal\c{c}ada: \url{https://www.eso.org/public/images/eso0831a/.}} \label{plotcs}
\end{figure}
Furthermore, the important case with $\alpha=1$ is realized when
$\xi\to \infty$, or similarly when $\Omega (\phi)\ll \frac{3}{2}
\frac{\left(\frac{d \Omega (\phi)}{d \phi}\right)^2}{\omega(\phi)}
$. The $R^p$ attractors yield a viable inflationary phenomenology,
see Ref. \cite{rpattractors}, with the spectral index of the
primordial scalar perturbations as a function of the canonical
scalar field being,
\begin{align}\label{spectralnsexplicit}
& n_s=\Big{(}\left(3 \alpha +(3 \alpha -2) p^2+(8-6 \alpha )
p-8\right) e^{2 \sqrt{\frac{2}{3}} \sqrt{\frac{1}{\alpha }} \kappa
\varphi }\\ \notag &-2 (p-1) (-3 \alpha +(3 \alpha -2) p+8)
e^{\sqrt{\frac{2}{3}} \sqrt{\frac{1}{\alpha }} \kappa  \varphi
}+(3 \alpha -8) (p-1)^2\Big{)}\\ \notag & \times 3 \alpha (p-1)^2
\left(e^{\sqrt{\frac{2}{3}} \sqrt{\frac{1}{\alpha }} \kappa
\varphi }-1\right)^2\, ,
\end{align}
and the tensor-to-scalar ratio is,
\begin{equation}\label{tensotoscalarasfunctionofscalarfield}
r=\frac{16 \left((p-2) e^{\sqrt{\frac{2}{3}} \sqrt{\frac{1}{\alpha
}} \kappa  \varphi }-2 p+2\right)^2}{3 \alpha  (p-1)^2
\left(e^{\sqrt{\frac{2}{3}} \sqrt{\frac{1}{\alpha }} \kappa
\varphi }-1\right)^2}\, .
\end{equation}
Also the free parameter $V_0$ of the potential is constrained to
have values
\begin{equation}\label{tilde}
V_s\sim 9.6\times 10^{-11}\, ,
\end{equation}
a results which originates from the constraints of the Planck data
on the Einstein frame amplitude $\Delta_s^2$ of the scalar
perturbations,
\begin{equation}\label{scalaramp}
\Delta_s^2=\frac{1}{24\pi^2}\frac{V(\varphi_f)}{M_p^4}\frac{1}{\epsilon(\varphi_f)}\,
.
\end{equation}
For the purposes of this paper, we shall consider several limiting
cases for the values of the parameter $\alpha$, mainly the cases
$\alpha\neq 1$, and the case $\alpha=1$, which corresponds to the
strong $\xi$ coupling theory. Also in order to have a viable
inflationary phenomenology, the parameter $p$ which is the
exponent in the $R^p$ attractors potential, has to take values in
the range $1.91\leq p \leq 1.99$. It proves that this is
irrelevant for NS studies, so we shall assume that $p=1.91$
without loss of generality. In the next section we shall specify
the values of the various functions involved in the TOV equations
of NS.

\section{Neutron Stars with $R^p$ Attractors}

For the purpose of studying NS in Einstein frame, we shall use the
Geometrized physical units system $G=c=1$, and we shall adopt the
notation of Ref. \cite{Pani:2014jra}.

The Jordan frame scalar-tensor theory has the following form,
\begin{equation}\label{ta}
\mathcal{S}=\int
d^4x\frac{\sqrt{-g}}{16\pi}\Big{[}\Omega(\phi)R-\frac{1}{2}g^{\mu
\nu}\partial_{\mu}\phi\partial_{\nu}\phi-U(\phi)\Big{]}+S_m(\psi_m,g_{\mu
\nu})\, ,
\end{equation}
and by performing the following conformal transformation,
\begin{equation}\label{ta1higgs}
\tilde{g}_{\mu \nu}=A^{-2}g_{\mu \nu}\,
,\,\,\,A(\phi)=\Omega^{-1/2}(\phi)\, ,
\end{equation}
we obtain the Einstein frame action,
\begin{equation}\label{ta5higgs}
\mathcal{S}=\int
d^4x\sqrt{-\tilde{g}}\Big{(}\frac{\tilde{R}}{16\pi}-\frac{1}{2}
\tilde{g}_{\mu \nu}\partial^{\mu}\varphi
\partial^{\nu}\varphi-\frac{V(\varphi)}{16\pi}\Big{)}+S_m(\psi_m,A^2(\varphi)g_{\mu
\nu})\, ,
\end{equation}
with $\varphi$ denoting the Einstein frame canonical scalar field
as in the previous section, and
\begin{equation}\label{potentialns1}
V(\varphi)=\frac{U(\phi)}{\Omega^2}\, .
\end{equation}
For the $R^p$ attractors with general $\alpha$, the important
function $A(\varphi)$ has the following form,
\begin{equation}\label{inducedA}
 A(\varphi)=e^{-\frac{1}{2}\sqrt{\frac{2}{3\alpha }}\varphi}\, ,
\end{equation}
therefore, the function $\alpha (\phi)$ which is defined as
follows,
\begin{equation}\label{alphaofvarphigeneraldef}
\alpha(\varphi)=\frac{d \ln A(\varphi)}{d \varphi}\, ,
\end{equation}
takes the form,
\begin{equation}\label{alphaofphifinalintermsofvarphi}
a(\varphi)=-\frac{1}{2}\sqrt{\frac{2}{3 \alpha }}\, .
\end{equation}
\begin{table}
  \begin{center}
    \caption{\emph{\textbf{CSI vs the $R^p$ Attractors for the SLy, APR and WFF1 EoSs for NS Masses $M\sim 2M_{\odot}$}}}
    \label{table1}
    \begin{tabular}{|r|r|r|r|}
     \hline
      \textbf{$R^p$ Attractor  Model}   & \textbf{APR} & \textbf{SLy} & \textbf{WFF1}
      \\  \hline
      \textbf{$\alpha=1$} & $M= 2.00\,M_{\odot}$ & $M= 2.01\, M_{\odot}$ & $M= 0.31\,
M_{\odot}$
\\  \hline
       \textbf{$\alpha=1$} & $R= 11.10$km & $R= 11.17$km
      &$R= 11.06$km
      \\  \hline
      \textbf{$\alpha=0.1$} & $M= 2.02\, M_{\odot}$ & $M= 2.00\, M_{\odot}$ & $M= 2.00\, M_{\odot}$
      \\  \hline
       \textbf{$\alpha=0.1$} & $R= 11.52$km &
$R= 11.818$km
      & $R= 11.012$km \\  \hline
      \textbf{$\alpha=8$} & $M= 2.00\, M_{\odot}$ & $M= 2.09\, M_{\odot}$ & $M= 0.32\, M_{\odot}$
      \\  \hline
       \textbf{$\alpha=8$} & $R= 11.08$km &
$R= 10.983$km
      & $R= 11.114$km \\  \hline
    \end{tabular}
  \end{center}
\end{table}
Finally, the Einstein frame scalar potential is given in Eq.
(\ref{rpinfinialpha}), which we also quote it here for reading
convenience,
\begin{equation}\label{rpinfinialphaupdate}
V(\varphi)= V_0\,e^{-2\sqrt{\frac{2}{3 \alpha}}
\varphi}\left(e^{\sqrt{\frac{2}{3\alpha }} \varphi}
-1\right)^{\frac{p}{p-1}}\, ,
\end{equation}
and in Geometrized units, the constraint on $V_0$ given in Eq.
(\ref{tilde}) becomes,
\begin{equation}\label{constraintonv0}
V_0\simeq 7.62\times 10^{-12}\, .
\end{equation}
For the study of NS physics, we shall consider the following
spherically symmetric metric,
\begin{equation}\label{tov1}
ds^2=-e^{\nu(r)}dt^2+\frac{dr^2}{1-\frac{2
m(r)}{r}}+r^2(d\theta^2+\sin^2\theta d\phi^2)\, ,
\end{equation}
which describes a static NS, where the function $m(r)$ describes
the total gravitational mass of the NS and $r$ stands for the
circumferential radius. In the following, we shall calculate
numerically the functions $\nu(r)$ and $\frac{1}{1-\frac{2
m(r)}{r}}$ following a simple procedure, in which the central
value of $\nu(r)$ and of the scalar field will be arbitrary and
will be optimally calculated numerically by using a double
shooting method.

The double shooting aims to find the optimal values of the central
values of $\nu(r)$ and of the scalar field, which guarantee that
the metric at numerical infinity becomes identical to the
Schwarzschild metric.
\begin{table}
  \begin{center}
    \caption{\emph{\textbf{CSI vs the $R^p$ Attractors for the SLy, APR and WFF1 EoSs for NS Masses $M\sim 1.4M_{\odot}$}}}
    \label{table2}
    \begin{tabular}{|r|r|r|r|}
     \hline
      \textbf{$R^p$ Attractors  Model}   & \textbf{APR} & \textbf{SLy} & \textbf{WFF1}
      \\  \hline
      \textbf{$\alpha=1$} & $M= 0.58\,M_{\odot}$ & $M= 1.41\, M_{\odot}$ & $M= 0.25\,
M_{\odot}$
\\  \hline
       \textbf{$\alpha=1$} & $R= 11.48$km & $R= 11.74$km
      &$R=  11.89$km
      \\  \hline
      \textbf{$\alpha=0.1$} & $M= 1.39\, M_{\odot}$ & $M= 1.39\, M_{\odot}$ & $M= 0.07\, M_{\odot}$
      \\  \hline
       \textbf{$\alpha=0.1$} & $R= 11.55$km &
$R= 12.04$km
      & $R= 11.79$km \\  \hline
      \textbf{$\alpha=8$} & $M= 0.64\, M_{\odot}$ & $M= 1.42\, M_{\odot}$ & $M= 0.28\, M_{\odot}$
      \\  \hline
       \textbf{$\alpha=8$} & $R= 11.45$km &
$R= 11.73$km
      & $R= 11.46$km \\  \hline
    \end{tabular}
  \end{center}
\end{table}
This procedure is different compared to standard General
Relativity (GR) NS, because in GR, the metric at the surface of
the star abruptly becomes the Schwarzschild metric. This is not
true in the scalar-tensor theories, because the scalar potential
and the non-minimally coupling function $A(\varphi)$ have
non-trivial effects on the NS beyond the surface of the star
(scalarization). The Einstein frame TOV equations take the
following form,
\begin{equation}\label{tov2}
\frac{d m}{dr}=4\pi r^2
A^4(\varphi)\varepsilon+\frac{r}{2}(r-2m(r))\omega^2+4\pi
r^2V(\varphi)\, ,
\end{equation}
\begin{equation}\label{tov3}
\frac{d\nu}{dr}=r\omega^2+\frac{2}{r(r-2m(r))}\Big{[}4\pi
A^4(\varphi)r^3P-4\pi V(\varphi)
r^3\Big{]}+\frac{2m(r)}{r(r-2m(r))}\, ,
\end{equation}
\begin{align}\label{tov4}
& \frac{d\omega}{dr}=\frac{4\pi r
A^4(\varphi)}{r-2m(r)}\Big{(}\alpha(\varphi)(\epsilon-3P)+
r\omega(\epsilon-P)\Big{)}-\frac{2\omega (r-m(r))}{r(r-2m(r))}\\
\notag &+\frac{8\pi \omega r^2 V(\varphi)+r\frac{d V(\varphi)}{d
\varphi}}{r-2 m(r)}\, ,
\end{align}
\begin{equation}\label{tov5}
\frac{dP}{dr}=-(\epsilon+P)\Big{[}\frac{1}{2}\frac{d\nu}{dr}+\alpha
(\varphi)\omega\Big{]}\, ,
\end{equation}
\begin{equation}\label{tov5newfinal}
\omega=\frac{d \varphi}{dr}\, ,
\end{equation}
with $\alpha (\varphi)$ being defined in Eq.
(\ref{alphaofvarphigeneraldef}). Also note that the energy density
$\epsilon$ and the pressure $P$ of the matter fluid are Jordan
frame quantities. We shall solve the TOV equations for both the
interior and the exterior of the NS, with the following set of
initial conditions being used,
\begin{equation}\label{tov8}
P(0)=P_c\, ,\,\,\,m(0)=0\, , \,\,\,\nu(0)\, ,=-\nu_c\, ,
\,\,\,\varphi(0)=\varphi_c\, ,\,\,\, \omega (0)=0\, .
\end{equation}
Both $\nu_c$ and $\varphi_c$ will be determined using a double
shooting method, and the numerical analysis shall be performed for
three distinct piecewise polytropic EoS, with the central part
being described by the SLy, WFF1 or the APR EoS.
\begin{figure}
\centering
\includegraphics[width=20pc]{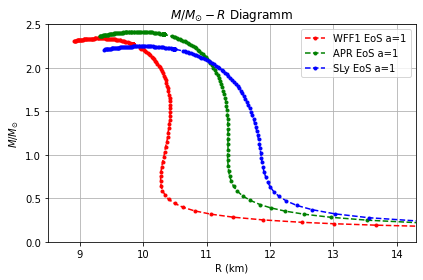}
\caption{The $M-R$ graphs for the $R^p$ attractor model for the
WFF1, APR and SLy EoSs, for $\alpha=1$} \label{quadplot1}
\end{figure}
For the calculation of the ADM mass in the Jordan frame we shall
use the following definition
\cite{Odintsov:2021qbq,Odintsov:2021nqa,Oikonomou:2021iid},
\begin{equation}\label{jordanframeADMmassfinal}
M=A(\varphi(r_E))\left(M_E-\frac{r_E^{2}}{2}\alpha
(\varphi(r_E))\frac{d\varphi
}{dr}\left(2+\alpha(\varphi(r_E))r_E\frac{d \varphi}{dr}
\right)\left(1-\frac{2 M_E}{r_E} \right) \right)\, .
\end{equation}
where $r_E$ denotes the Einstein frame circumferential radius of
the NS, and also we define  $\frac{d\varphi }{dr}=\frac{d\varphi
}{dr}\Big{|}_{r=r_E}$. Finally, the circumferential radii of the
NS in the Jordan and Einstein frames are related as
$R=A(\varphi(R_s))\, R_s$. We shall measure the Jordan frame mass
in solar masses $M_{\odot}$ and the Jordan frame radius in
kilometers.

\subsection{Results of the Numerical Analysis}

Let us now present the results of our numerical analysis on the NS
phenomenology of the $R^p$ attractors. We considered three
characteristic cases of attractors, corresponding to three values
of $\alpha$, namely $\alpha=1$, $\alpha=0.1$ and $\alpha=8$.
\begin{figure}
\centering
\includegraphics[width=20pc]{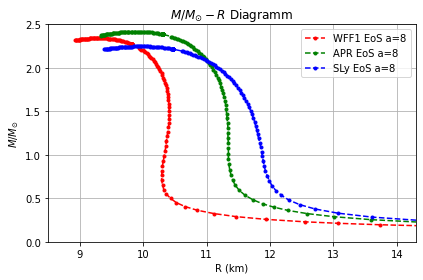}
\caption{The $M-R$ graphs for the $R^p$ attractor model for the
WFF1, APR and SLy EoSs, for $\alpha=8$.} \label{indplot1}
\end{figure}
All these values of $\alpha$ produce a viable inflationary
phenomenology as was shown in Ref. \cite{rpattractors}. Here we
shall present the $M-R$ graphs for the $R^p$ attractors for the
three values of $\alpha$. Accordingly the results will be
confronted with three distinct constraints on NS radii for
specific mass NS. Specifically we shall use the following
constraints, developed in Refs. \cite{Altiparmak:2022bke},
\cite{Raaijmakers:2021uju} and \cite{Bauswein:2017vtn} to which we
shall refer to as CSI, CSII and CSIII respectively. The CSI
indicates that the radius of an $1.4M_{\odot}$ mass NS should be
$R_{1.4M_{\odot}}=12.42^{+0.52}_{-0.99}$ and furthermore, the
radius of an $2M_{\odot}$ mass NS should be
$R_{2M_{\odot}}=12.11^{+1.11}_{-1.23}\,$km. Accordingly, CSII
indicates that the radius of an $1.4M_{\odot}$ mass NS should be
$R_{1.4M_{\odot}}=12.33^{+0.76}_{-0.81}\,\mathrm{km}$. Lastly,
CSIII indicates that the radius of an $1.6M_{\odot}$ mass NS
should be larger than $R_{1.6M_{\odot}}=12.42^{+0.52}_{-0.99}\,$km
and the radius of a NS with maximum mass should be larger than
$R_{M_{max}}>10.68^{+0.15}_{-0.04}\,$km.
\begin{figure}
\centering
\includegraphics[width=20pc]{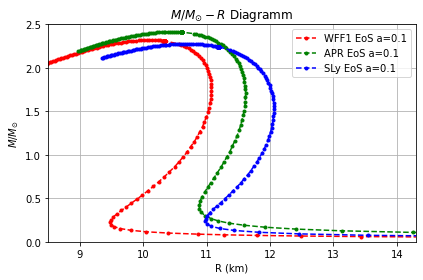}
\caption{The $M-R$ graphs for the $R^p$ attractor model for the
WFF1, APR and SLy EoSs, for $\alpha=0.1$.} \label{quadplot2}
\end{figure}
The constraints CSI, CSII and CSIII are pictorially represented in
Fig. \ref{plotcs}\footnote{This media was originally created by
the European Southern Observatory (ESO). I edited the figure for
demonstrative purposes. Their website states: ''Unless
specifically noted, the images, videos, and music distributed on
the public ESO website, along with the texts of press releases,
announcements, pictures of the week, blog posts and captions, are
licensed under a Creative Commons Attribution 4.0 International
License, and may on a non-exclusive basis be reproduced without
fee provided the credit is clear and visible.''}. Using a double
shooting LSODA python 3 numerical integration method
\cite{niksterg}, and also by setting the numerical infinity at
$r\sim 67.943$ km, at this point we shall present our results,
which can be seen in the $M-R$ plots and the tables appearing in
this work. Note that the numerical infinity plays an important
role for the double shooting method, in order for the scalar field
effects to be switched off at the numerical infinity.

\begin{figure}
\centering
\includegraphics[width=20pc]{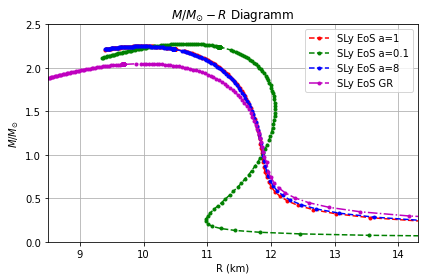}
\caption{The $M-R$ graphs of the $R^p$ attractors for $\alpha=1$
(red curve), $\alpha=0.1$ (green curve), $\alpha=8$ (blue curve)
and the GR (magenta curve) for the SLy EoS .} \label{compplot2}
\end{figure}

\begin{figure}
\centering
\includegraphics[width=20pc]{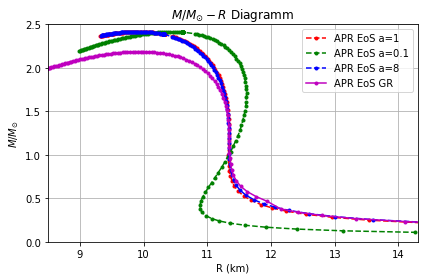}
\caption{The $M-R$ graphs of the $R^p$ attractors for $\alpha=1$
(red curve), $\alpha=0.1$ (green curve), $\alpha=8$ (blue curve)
and the GR (magenta curve) for the APR EoS.} \label{compplot3}
\end{figure}

To start with, in Figs. \ref{quadplot1}, \ref{quadplot2} and
\ref{indplot1} we present the $M-R$ graphs of the $R^p$ attractors
for $\alpha=1$, $\alpha=0.1$ and $\alpha=8$ NS respectively, for
the WFF1 EoS (red curve), the APR EoS (green curve) and the SLy
EoS (blue curve). In all the cases, the maximum masses of the NS
are larger compared to the GR case. Also it is notable that the
$\alpha=1$ case is quite similar to the  $\alpha=8$ case, however
strong differences are observed for the $\alpha=0.1$ case.
\begin{figure}
\centering
\includegraphics[width=20pc]{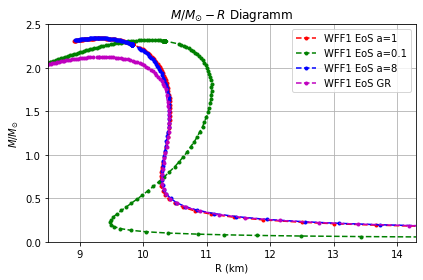}
\caption{The $M-R$ graphs of the $R^p$ attractors for $\alpha=1$
(red curve), $\alpha=0.1$ (green curve), $\alpha=8$ (blue curve)
and the GR (magenta curve) for the WFF1 EoS.} \label{compplot1}
\end{figure}
Also in Figs. \ref{compplot1}, \ref{compplot2} and \ref{compplot3}
we present for each EoS the $M-R$ graphs of the $R^p$ attractors
for $\alpha=1$ (red curves), $\alpha=0.1$ (green curves),
$\alpha=8$ (blue curves) and the GR (magenta curves) for the WFF1
EoS (upper left plot) the SLy EoS (upper right) and the APR EoS
(bottom plot). Now let us present the confrontation of the $R^p$
attractor NS with the constraints CSI, CSII and CSIII.

The results of our analysis regarding the confrontation of the
$R^p$ inflationary attractors models with the observational
constraints on NS, namely CSI, CSII, AND CSIII are presented in
Tables \ref{table1}-\ref{table5}. For the case with $\alpha=1$,
the SLy EoS is compatible with all the constraints, with regard to
the APR, it is not compatible with CSII, the first constraint of
CSI, but it is compatible with the second constraint of CSII and
the CSIII constraints. Also the WFF1 case is incompatible with all
the constraints. For the case with $\alpha=0.1$, the SLy EoS is
compatible with all the constraints, and interestingly enough, for
this case the APR is also compatible with all the constraints.
However, in this case the WFF1 EoS satisfies the second constraint
of CSI and also satisfies all the constraints of CSIII.
Finally, for the case with $\alpha=1$, the SLy EoS is compatible
with all the constraints, with regard to the APR, it is not
compatible with CSII, and the first constraint of CSI, but it is
compatible with the second constraint of CSII and the CSIII
constraints.

Also the WFF1 case is incompatible with all the constraints, save
the first constraint of CSIII. Hence, the viable NS
phenomenologies that pass all the tests imposed by the constraints
CSI, CSII and CSIII, are provided by all the SLy cases for all the
values of the parameter $\alpha$, and also by the APR EoS, only
when $\alpha=0.1$. Thus apparently, obtaining a viable NS
phenomenology nowadays is not as easy it was before the GW170817
event. Also regarding the $R^p$ attractors, these can be
discriminated in NS, for different values of $\alpha$, especially
for $0.1<\alpha<1$. However, as $\alpha$ grows larger than unity,
it seems that $R^p$ attractors provide an almost identical NS
phenomenology. This is a notable feature for the class of $R^p$
attractors.
\begin{table}
  \begin{center}
    \caption{\emph{\textbf{CSIII vs the $R^p$ Attractors for the SLy, APR and WFF1 EoSs for NS Masses $M\sim 1.6M_{\odot}$}}}
    \label{table4}
    \begin{tabular}{|r|r|r|r|}
     \hline
      \textbf{$R^p$ Attractors Model}   & \textbf{APR} & \textbf{SLy} & \textbf{WFF1}
      \\  \hline
      \textbf{$\alpha=1$} & $M= 1.60\,M_{\odot}$ & $M= 1.60\, M_{\odot}$ & $M= 1.61\,
M_{\odot}$
\\  \hline
       \textbf{$\alpha=1$} & $R= 11.30$km & $R= 11.63$km
      &$R= 10.41$km
      \\  \hline
      \textbf{$\alpha=0.1$} & $M= 1.61\, M_{\odot}$ & $M= 1.60\, M_{\odot}$ & $M= 1.59\, M_{\odot}$
      \\  \hline
       \textbf{$\alpha=0.1$} & $R= 11.61$km &
$R= 12.05$km
      & $R= 11.05$km \\  \hline
      \textbf{$\alpha=8$} & $M= 1.61\, M_{\odot}$ & $M= 1.60\, M_{\odot}$ & $M= 1.58\, M_{\odot}$
      \\  \hline
       \textbf{$\alpha=8$} & $R= 11.28$km &
$R= 12.05$km
      & $R= 10.40$km \\  \hline
    \end{tabular}
  \end{center}
\end{table}
Before closing, we need to discuss an important issue, having to
do with the NS phenomenology of inflationary potentials, with
regard to the tidal deformability of NSs, the radial perturbations
of static NSs and finally the overall stability of NSs, by also
taking into account the constraints imposed by the GW170817 event.
This issue however extends further from the aims and scopes of
this article, since a whole article could be devoted to these
issues, see for example Refs. \cite{Brown:2022kbw} and
\cite{Yang:2022ees}, in which these issues are addressed in the
context of scalar-tensor gravity \cite{Brown:2022kbw} and in
unimodular gravity \cite{Yang:2022ees}.

\section*{Concluding Remarks}

In this article we studied the NS phenomenology of the $R^p$
inflationary attractor scalar-tensor models in the Einstein frame.
The $R^p$ attractors constitute a class of models in the Einstein
frame, which originate from a large number of different models in
the Jordan frame These distinct Jordan frame models result to the
same phenomenology in the Einstein frame and this feature
justifies the terminology inflationary attractors. Our aim was to
investigate whether these attractor models can be distinguished
when NSs are considered.
As we showed the NS phenomenology corresponding to different
values of the parameter $\alpha$ which characterizes the
attractors, is in general different for $\alpha<1$, however the
models for $\alpha>1$ show many similarities and generate almost
identical $M-R$ diagrams. We also confronted the NS phenomenology
of the $R^p$ attractors to several NS constraints, which we named
CSI, CSII and CSIII. The constraint CSI was developed in Ref.
\cite{Altiparmak:2022bke} and indicates that the radius of an
$1.4M_{\odot}$ mass NS has to be
$R_{1.4M_{\odot}}=12.42^{+0.52}_{-0.99}$ while the radius of an
$2M_{\odot}$ mass NS has to be
$R_{2M_{\odot}}=12.11^{+1.11}_{-1.23}\,$km. The constraint CSII
was developed in Ref. \cite{Raaijmakers:2021uju} and indicates
that the radius of an $1.4M_{\odot}$ mass NS has to be
$R_{1.4M_{\odot}}=12.33^{+0.76}_{-0.81}\,\mathrm{km}$ and the
constraint CSIII was developed in Ref. \cite{Bauswein:2017vtn} and
indicates that the radius of an $1.6M_{\odot}$ mass NS has to be
larger than $R_{1.6M_{\odot}}=12.42^{+0.52}_{-0.99}\,$km while the
radius of the maximum mass NS has to be larger than
$R_{M_{max}}>10.68^{+0.15}_{-0.04}\,$km. Our analysis indicated
that for $R^p$ attractors, for the case with $\alpha=1$, only the
SLy EoS is compatible with all the constraints, while the APR is
not compatible with CSII, the first constraint of CSI, but it is
compatible with the second constraint of CSII and the CSIII
constraints. Also the WFF1 case is incompatible with all the
constraints.
\begin{table}
  \begin{center}
    \caption{\emph{\textbf{CSII vs the $R^p$ Attractors for the SLy, APR and WFF1 EoSs for NS Masses $M\sim 1.4M_{\odot}$}}}
    \label{table3}
    \begin{tabular}{|r|r|r|r|}
     \hline
      \textbf{$R^p$ Attractors Model}   & \textbf{APR} & \textbf{SLy} & \textbf{WFF1}
      \\  \hline
      \textbf{$\alpha=1$} & $M= 0.52\,M_{\odot}$ & $M= 1.41\, M_{\odot}$ & $M= 0.25\,
M_{\odot}$
\\  \hline
       \textbf{$\alpha=1$} & $R= 11.56$km & $R= 11.74$km
      &$R= 11.89$km
      \\  \hline
      \textbf{$\alpha=0.1$} & $M= 1.39\, M_{\odot}$ & $M= 1.39\, M_{\odot}$ & $M= 0.07\, M_{\odot}$
      \\  \hline
       \textbf{$\alpha=0.1$} & $R= 11.55$km &
$R= 12.04$km
      & $R= 11.79$km \\  \hline
      \textbf{$\alpha=8$} & $M= 0.53\, M_{\odot}$ & $M= 1.42\, M_{\odot}$ & $M= 0.25\, M_{\odot}$
      \\  \hline
       \textbf{$\alpha=8$} & $R= 11.60$km &
$R= 11.738$km
      & $R= 11.944$km \\  \hline
    \end{tabular}
  \end{center}
\end{table}
\begin{table}
  \begin{center}
    \caption{\emph{\textbf{CSIII vs the $R^p$ Attractors for the SLy, APR and WFF1 EoSs for Maximum NS Masses}}}
    \label{table5}
    \begin{tabular}{|r|r|r|r|}
     \hline
      \textbf{$R^p$ Attractors  Model}   & \textbf{APR} & \textbf{SLy} & \textbf{WFF1}
      \\  \hline
      \textbf{$\alpha=1$} & $M=  2.41\,M_{\odot}$ & $M= 2.24\, M_{\odot}$ & $M= 2.33\,
M_{\odot}$
\\  \hline
       \textbf{$\alpha=1$} & $R= 9.91$km & $R= 9.99$km
      &$R= 9.30$km
      \\  \hline
      \textbf{$\alpha=0.1$} & $M= 2.41\, M_{\odot}$ & $M= 2.27\, M_{\odot}$ & $M= 2.32\, M_{\odot}$
      \\  \hline
       \textbf{$\alpha=0.1$} & $R= 10.40$km &
$R= 10.09$km
      & $R= 11.06$km \\  \hline
      \textbf{$\alpha=8$} & $M= 2.41\, M_{\odot}$ & $M= 2.27\, M_{\odot}$ & $M= 2.34\, M_{\odot}$
      \\  \hline
       \textbf{$\alpha=8$} & $R= 9.91$km &
$R= 10.72$km
      & $R= 9.28$km \\  \hline
    \end{tabular}
  \end{center}
\end{table}

For the case with $\alpha=0.1$, which is the most interesting case
phenomenologically, the SLy EoS is compatible with all the
constraints, and for this case the APR is also compatible with all
the constraints. However, in this case the WFF1 EoS satisfies the
second constraint of CSI and also satisfies all the constraints of
CSIII. Finally, for the case with $\alpha=1$, only the SLy EoS is
compatible with all the constraints while the APR is not
compatible with CSII, and the first constraint of CSI, but it is
compatible with the second constraint of CSII and the CSIII
constraints. Finally, the WFF1 case is incompatible with all the
constraints, save the first constraint of CSIII. Our results
indicate two main research lines, firstly that NS phenomenology
for scalar-tensor theories is not easily rendered viable, since a
large number of astrophysical and cosmological constraints have to
be satisfied in order for the viability of the model to be
guaranteed. Thus a simple parameter assigning is  not the correct
way to study NS nowadays, both cosmology and astrophysics
constrain in a rigid way NSs. Secondly, several inflationary
attractors which are indistinguishable at the cosmological level,
may be discriminated to some extent when their NS phenomenology is
considered. This research line is not the general rule though, so
work is in progress toward comparing a large sample of
cosmological attractors with respect to their NS phenomenology.
Finally, let us note that the scalar-tensor inflationary framework
we used in this work cannot be considered more advantageous
compared to other modified gravity theories, it is one of the many
possible modified gravity descriptions of the nature of NSs.

\section*{Acknowledgments}

This work was supported by MINECO (Spain), project
PID2019-104397GB-I00 (S.D.O). This work by S.D.O was also
partially supported by the program Unidad de Excelencia Maria de
Maeztu CEX2020-001058-M, Spain.

\textbf{Data availability.} No new data were generated or analysed
in support of this research.

\bibliographystyle{mnras}

\label{lastpage}
\end{document}